\documentclass[]{seismica}

\usepackage[english]{babel}
\title{
	Integrating {$\boldsymbol{b}$}-Value and Background Seismicity Rate for Spatial Earthquake Forecasting in the Alborz Region, Northern Iran}

\author[1]{Muhammed Hossein Mousavi
	\orcid{0009-0003-2282-8757}
}
\author[2,3]{Hamzeh Mohammadigheymasi
	\orcid{0000-0002-1136-2651}
}
\author[4]{Parva Sadeghi Alavijeh
}
\author[5]{Marjan Tourani
}

\affil[1]{Department of Physics Education, Farhangian University, Tehran, Iran}

\affil[2]{Atmosphere and Ocean Research Institute (AORI), The University of Tokyo, Kashiwa, Japan}

\affil[3]{Department of Computer Sciences, University of Beira Interior, Covilha, Portugal}

\affil[4]{Department of Physics, University of Kashan, Kashan, Iran}

\affil[5]{Department of Geological Engineering, Ankara University, Tectonics Research Group, Ankara, Türkiye}

\begin{document}
\nolinenumbers


\makeseistitle{
	\begin{summary}{Abstract}
In this study, we evaluate the spatial forecasting skill of the $b$-value and background seismicity rate $\mu$ across the Alborz region using a homogenized catalog of 23,961 earthquakes ($M \geq 1.5$) recorded by the Iranian Seismological Center between 2006 and 2024. Forecast performance for $M \geq 4.0$ and $M \geq 4.5$ is assessed using Molchan error diagrams, probability gain, probability difference, and the modified area skill score. The results show that $\mu$ provides a consistently strong spatial signal, with Molchan curves well below the random baseline and probability gains of 5--6 at low alarm rates, reflecting the persistent clustering of seismicity along major Alborz faults. The $b$-value exhibits limited skill at lower magnitudes but improves steadily with increasing magnitude; its skill score becomes positive above $M \approx 5.3$, indicating that $b$-value anomalies begin to capture meaningful stress concentrations only for larger events. Spatial patterns reveal low $b$ zones along active reverse and strike-slip structures and high $\mu$ zones following long-term seismicity clusters, underscoring their complementary physical roles. Retrospective testing confirms this complementarity: the combined $b$--$\mu$ forecast achieves detection rates of 0.81--0.83 at spatial alarm rates of 0.43 and 0.36 for $M \geq 4.0$ and $M \geq 4.5$, respectively, representing the most efficient forecast configuration among all tested models. These findings demonstrate that integrating stress-state and tectonic-loading indicators yields a more efficient and physically grounded framework for operational earthquake forecasting in the Alborz region.
	\end{summary}
	\begin{summary}{Non-technical summary}
Earthquakes are a major hazard in northern Iran's active Alborz region. To improve forecasting, we analyzed nearly 24,000 earthquakes recorded between 2006 and 2024, evaluating two geological indicators: the background seismicity rate, which tracks tectonic loading, and the b value, representing deep crustal stress. We found that the background seismicity rate is a highly stable spatial predictor across all magnitudes, as earthquakes naturally cluster along active faults. In contrast, the stress sensitive b value is less informative for minor tremors but becomes a powerful forecasting tool for larger, potentially damaging earthquakes (magnitude 5.3 and above) as faults approach their breaking point. Because these indicators track complementary physical processes, representing ongoing tectonic loading versus current structural pressure, combining them into a unified forecast outperforms either tool used alone. This integrated model successfully anticipates over 80\% of moderate to large earthquakes while minimizing false alarms. Our study offers a testable forward looking hazard map for 2025 to 2029, establishing a stronger, physically grounded foundation for disaster mitigation and seismic risk reduction in active mountain belts.
	\end{summary}
}
	

\section{Introduction}
\label{sec:intro}
Earthquakes are among the most destructive natural hazards on Earth, capable of collapsing buildings, severing lifelines, and disrupting the social and economic fabric of entire communities for years or decades after the event \cite{kazama2012, cui2011}. The 2004 {$M_w$}~9.0 Sumatra, 2008 {$M_w$}~8.0 Wenchuan, and 2010 {$M_w$}~7.0 Haiti earthquakes each illustrated, across very different tectonic and societal settings, how comprehensively a single rupture can reshape the trajectory of entire nations \cite{nanjo2012, shi2018, cavallo2010}. The desire to anticipate such events before they occur has motivated earthquake forecasting research for much of the past century, even as that research has required, as \cite{jordan2010} and \cite{geller1997} have frankly acknowledged, repeated reassessment of what the field can and cannot deliver. The methods brought to bear on this problem have been remarkably diverse. Seismic gap analysis, characteristic earthquake models, static and dynamic stress transfer, aftershock sequence modeling, pattern recognition, and more recently machine learning have all contributed to our understanding of how and where earthquakes occur \cite{bakun1985, king1994, brodsky2006, keilisborok1990m8, wang2004}. Yet reliable intermediate- to long-term forecasting across tectonically complex regions remains a significant scientific challenge. Classical models were largely developed for the sparser catalogs of earlier monitoring eras and have not always kept pace with the richer datasets now available, while the fundamental rarity of large damaging earthquakes continues to limit the statistical power of any validation exercise \cite{jordan2010, michael2020, beroza2021}.
Against this background, the Gutenberg--Richter {$b$}-value has established itself as one of the most physically meaningful and operationally tractable indicators of seismic potential \cite{gutenberg1944}. Its utility for hazard assessment lies not simply in this descriptive role but in its sensitivity to crustal stress: both laboratory experiments and field studies have consistently shown that {$b$}-values decrease as differential stress increases and rise where the crust is more heterogeneous or heavily fractured \cite{scholz2015, beall2022}. This physical relationship gives spatial {$b$}-value variations genuine diagnostic power. Anomalously low {$b$}-values were documented in the epicentral regions of the 2004 Sumatra, 2008 Wenchuan, and 2011 Tohoku earthquakes prior to their occurrence \cite{nanjo2012, elisa2014, shi2018}, lending empirical support to the interpretation that {$b$}-value mapping can track where fault systems may be approaching failure. Natural time analyses have extended this picture further, connecting pre-mainshock {$b$}-value decreases to critical state dynamics in the evolving seismic system \cite{sarlis2015, nanjo2022, convertito2024}. One important caveat applies throughout: {$b$}-value estimates are sensitive to the magnitude of completeness ({$M_c$}) of the underlying catalog, and careful {$M_c$} assessment is a necessary prerequisite for drawing meaningful conclusions. 
The background seismicity rate ({$\mu$}) approaches the same problem from a complementary angle. By isolating the rate of independent earthquake occurrence after systematic removal of aftershocks, foreshocks, and swarms through declustering, {$\mu$} captures the component of seismicity most directly tied to ongoing tectonic stress accumulation \cite{zaliapin2013, vanstiphout2012}. The rate-state friction framework provides a clear physical basis for this connection: as fault systems are loaded tectonically, background seismicity rates respond accordingly \cite{hainzl2006, heimisson2018, nandan2017}. The observational record supports this interpretation. Large earthquakes tend to nucleate preferentially in regions of elevated background seismicity \cite{ogata2022}, and variations in {$\mu$} have been associated with independently measurable signals in crustal strain, transient stress fields, and subsurface fluid dynamics \cite{stevens2021}. What makes the joint use of these two parameters particularly attractive is that they are physically complementary rather than redundant. The {$b$}-value is primarily sensitive to the structural and stress-dependent organization of fault systems, while {$\mu$} reflects the ambient intensity of tectonic loading. Together, they capture dimensions of fault system behavior that neither reveals in isolation. Regions where low {$b$}-values coincide with high background seismicity rates are of particular concern, as they identify fault segments that are both highly stressed and actively loaded, conditions broadly associated with proximity to failure. Where high {$b$}-values and low background rates co-occur, the picture is one of more heterogeneous and less critically stressed crustal volumes \cite{ahadov2022, mousaviyan2025}. This joint framework provides a physically grounded basis for spatial seismic hazard assessment in settings where fault complexity limits what any single parameter can usefully reveal.
The Alborz Fold-Thrust Belt of northern Iran is one such setting. Shaped by the ongoing convergence of the Arabian and Eurasian plates at approximately 22~mm/yr, the belt accommodates deformation through a combination of reverse and strike-slip faulting along active structures that include the North Tehran, Mosha, and Khazar faults, among others \cite{berberian1999, tourani2024}. The seismic consequences of this tectonic configuration are well documented. Among the most destructive regional events are the 2004 {$M_w$}~6.2 Baladeh earthquake and the 2010 {$M_w$}~5.9 Kuh-Zar earthquake \cite{tatar2007, gheitanchi2005, shahvar2013}. Continued investment in the Iranian national seismic network has produced catalogs of sufficient density and resolution to support detailed statistical seismicity analysis. Nevertheless, the joint spatial predictive value of the {$b$}-value and background seismicity rate has not been systematically evaluated within this tectonic setting, and that gap is what the present study directly addresses. In this study, {$\mu$} refers to the background seismicity rate estimated after declustering within the Epidemic Type Aftershock Sequence (ETAS) framework. We map both parameters across the Alborz Fold-Thrust Belt, assess their individual and combined ability to forecast the spatial distribution of moderate to large earthquakes using the Molchan Error Diagram (MED), Probability Gain (PG), and Probability Difference (PD), and interpret the outcomes in the context of regional seismic hazard. The broader aim is to contribute to the methodological development of seismicity-based forecasting in tectonically complex environments where physically interpretable and statistically robust approaches are most needed.
The remainder of this paper is organized as follows. Section~\ref{sec:methodology} outlines the methodological framework and statistical procedures. Section~\ref{sec:data} describes the seismic dataset and catalog preparation. Section~\ref{sec:results} presents and discusses the forecasting results. Section~\ref{sec:conclusion} summarizes the principal findings and their implications for seismic hazard assessment in the Alborz region and comparable tectonic settings.

\section{Methodology}
\label{sec:methodology}

\subsection{{$b$}-Value Estimation and Spatial Mapping}
The frequency--magnitude distribution of earthquakes is described by the classical Gutenberg--Richter (G--R) relation \cite{gutenberg1944}:
\begin{equation}
\log_{10} N(M) = a - bM
\label{eq:gr_relation}
\end{equation}
where $N(M)$ is the cumulative number of earthquakes with magnitude greater than or equal to $M$, $a$ reflects the regional seismic productivity, and $b$ characterizes the relative occurrence of large to small events.

The {$b$}-value is estimated using the maximum likelihood method \cite{aki1965, wiemer1997}:
\begin{equation}
b = \frac{\log_{10} e}{\bar{M} - M_c}
\label{eq:mle_b}
\end{equation}
where $\bar{M}$ is the mean magnitude of the earthquakes in the sample, and $M_c$ is the magnitude of completeness. The associated standard error of the estimate is given by \cite{shibolt1982}:
\begin{equation}
\sigma_b = 2.30 b^2 \sqrt{\frac{\sum_{i=1}^n (M_i - \bar{M})^2}{n(n-1)}}
\label{eq:shi_bolt}
\end{equation}
The magnitude of completeness $M_c$ is determined by the Maximum Curvature method \cite{wiemer2000}, defined as the magnitude bin with the highest frequency in the non-cumulative frequency--magnitude distribution.

To resolve the spatial variability of the {$b$}-value, we adopt the Hierarchical Space--Time Point-Process Model (HIST-PPM) framework \cite{ogata2020manual}. Under this framework, $\beta = b \ln(10)$ is treated as a spatially continuous field parameterized through a two-dimensional spline representation:
\begin{equation}
\beta(x, y) = \sum_{j=1}^m \theta_j B_j(x, y)
\label{eq:spline_beta}
\end{equation}
where $\theta_j$ denotes the spline coefficients and $B_j(x, y)$ represent the piecewise linear basis functions. The study region is tessellated into Delaunay triangular elements centered on earthquake epicenters, adaptively concentrating spatial resolution where seismicity is densest. To prevent overfitting, $\theta$ is estimated by maximizing the penalized log-likelihood:
\begin{equation}
\ln L_P(\theta) = \ln L(\theta) - \omega \Phi(\theta)
\label{eq:penalized_likelihood}
\end{equation}
where $\ln L(\theta)$ is the point-process log-likelihood, $\Phi(\theta)$ is the spatial roughness penalty function, and $\omega$ is the smoothing hyperparameter. The roughness penalty weight $\omega$ is optimized by minimizing the Akaike Bayesian Information Criterion (ABIC):
\begin{equation}
\text{ABIC} = -2 \ln L_P(\hat{\theta}) + 2 \ln \det\left(\mathbf{H}_P(\hat{\theta})\right) - 2 \ln \det(\omega \mathbf{G}) + 2k
\label{eq:abic}
\end{equation}
where $\hat{\theta}$ is the maximum penalized likelihood estimate, $\mathbf{H}_P$ is the Hessian of the penalized log-likelihood, $\mathbf{G}$ is the roughness penalty matrix, and $k$ is the number of hyperparameters. Final {$b$}-values at epicenters are interpolated to a $0.1^\circ \times 0.1^\circ$ grid via linear interpolation within triangular elements.

\subsection{Background Seismicity Rate}
The spatially varying background seismicity rate $\mu(x, y)$ is estimated using the Epidemic-Type Aftershock Sequence (ETAS) model \cite{zhuang2002}, which decomposes the observed seismicity into a stationary background process and a triggered aftershock component. The total conditional seismic intensity $\lambda(t, x, y)$ is given by:
\begin{equation}
\lambda(t, x, y) = \mu(x, y) + \sum_{i: t_i < t} \kappa(M_i) g(t - t_i) f(x - x_i, y - y_i; M_i)
\label{eq:etas_intensity}
\end{equation}
The aftershock productivity follows $\kappa(M_i) = K_0 e^{\alpha(M_i - M_c)}$, and the temporal decay is governed by the modified Omori--Utsu law:
\begin{equation}
g(t) = \frac{p-1}{c} \left(1 + \frac{t}{c}\right)^{-p}
\label{eq:omori}
\end{equation}
All model parameters $(K_0, \alpha, c, p, d, q, \gamma)$ are estimated by maximizing the log-likelihood over the observation window.

For each earthquake, the probability of belonging to the background process (the declustering probability) is calculated as:
\begin{equation}
\rho_i = \frac{\mu(x_i, y_i)}{\lambda(t_i, x_i, y_i)}
\label{eq:declust_prob}
\end{equation}
The background seismicity rate is then reconstructed via a variable-bandwidth Gaussian kernel estimator:
\begin{equation}
\mu(x, y) = \frac{1}{T} \sum_{i=1}^N \frac{\rho_i}{2\pi d_i^2} \exp\left( -\frac{(x - x_i)^2 + (y - y_i)^2}{2 d_i^2} \right)
\label{eq:kernel_estimation}
\end{equation}
where $T$ is the total observation period and the bandwidth $d_i$ is the radius of the smallest circle centered at $(x_i, y_i)$ enclosing at least $n_p = 3$ neighboring events, subject to a minimum threshold $\varepsilon = 0.02$. Estimation proceeds iteratively: the updated $\mu(x, y)$ is re-injected into the ETAS model and parameters are re-estimated until convergence \cite{zhuang2002}.

\subsection{Evaluation of Forecast Performance}
Forecast performance is evaluated using the Molchan Error Diagram \cite{molchan1990}, which traces the relationship between the alarming rate ($\tau$), defined as the fraction of grid cells placed under alarm, and the missing rate ($1 - \nu$), defined as the fraction of target earthquakes occurring outside alarmed cells. For the {$b$}-value, cells with $b < b_{\text{thr}}$ are alarmed; for the background rate, cells with $\mu > \mu_{\text{thr}}$ are alarmed. A forecast outperforming random guessing lies below the diagonal in the error diagram.

Two scalar metrics further quantify forecast skill. The Probability Gain (PG) is defined as:
\begin{equation}
\text{PG} = \frac{\nu}{\tau}
\label{eq:probability_gain}
\end{equation}
which measures the improvement in detection relative to a random alarm of equal size ($\text{PG} > 1$ indicates skill). The Probability Difference (PD) is defined as:
\begin{equation}
\text{PD} = \nu - \tau
\label{eq:probability_difference}
\end{equation}
providing a linear measure of net forecast benefit, with $\text{PD} = 0$ for a random guess. The threshold corresponding to the maximum PD is adopted as the operational decision boundary, as it optimally balances detection gain against alarm cost.

To integrate both predictors, a joint alarm criterion is applied: a grid cell is flagged only when $b < b_{\text{thr}}$ and $\mu > \mu_{\text{thr}}$ are simultaneously satisfied. This compound strategy uses low-{$b$} anomalies to identify zones of high differential stress concentration, while high background seismicity rates serve as a secondary spatial filter to exclude aseismic or low-stress regions, effectively reducing false alarm rates and enhancing the physical interpretability of the spatial forecast.

\section{Seismic Dataset}
\label{sec:data}
\begin{figure}[htbp]
  \centering  \includegraphics[width=0.99\textwidth]{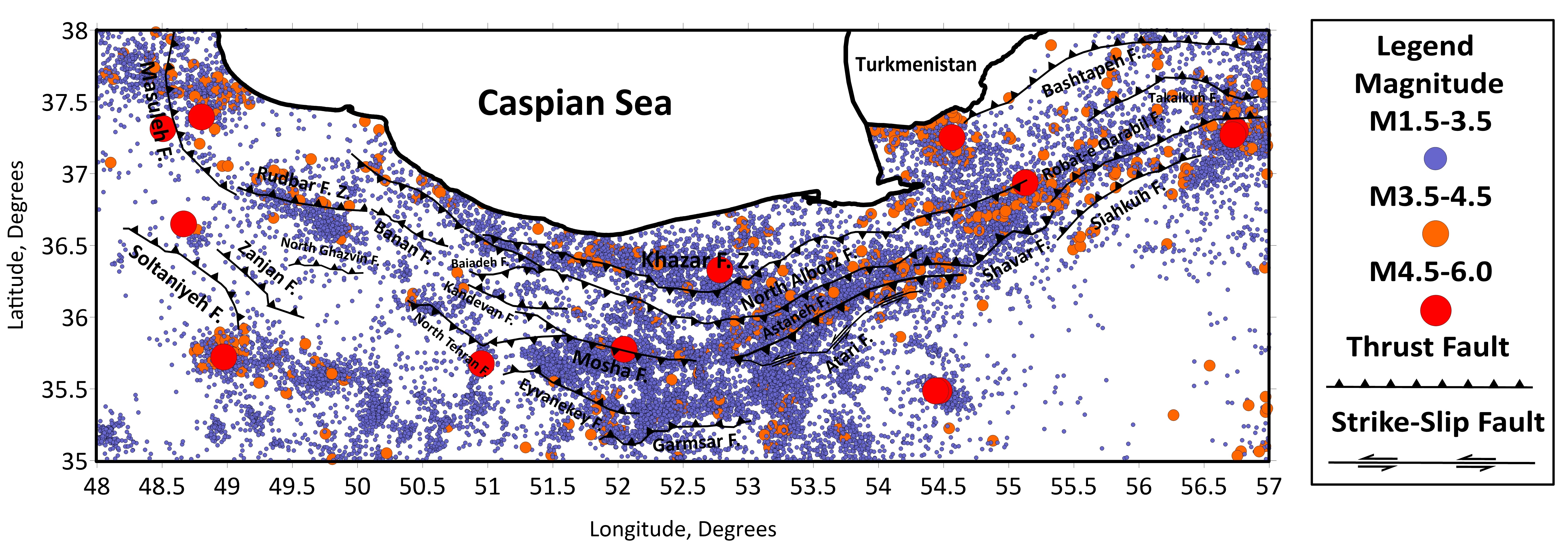}
  \caption{Seismicity map of the study region showing earthquake distribution and major active faults in the Alborz region.}
  \label{fig:seismicity_map}
\end{figure}

The earthquake catalog used in this study was compiled from the Iranian Seismological Center (IRSC), which routinely reports local events in terms of Nuttli magnitude ({$M_N$}) \cite{lamessa2019}. Because {$M_N$} is not directly comparable with the moment-magnitude ({$M_w$}) scales used in most regional and global studies, we homogenized the catalog by converting all {$M_N$} values to {$M_w$}. For this purpose, we applied the empirical linear relationship developed by \cite{karimiparidari2013}, which has been widely adopted for Iranian seismicity and provides a stable basis for magnitude unification. This step ensures internal consistency within the dataset and improves the reliability of subsequent seismicity-based analyses. 
The final catalog contains 23,961 earthquakes with magnitudes {$M \ge 1.5$}, spanning an 18-year period from 1 January 2006 to 31 October 2024. This long observational window captures both background seismicity and the moderate events characteristic of the Alborz region, providing a robust foundation for completeness assessment, spatial pattern analysis, and forecasting experiments. Figure~\ref{fig:seismicity_map} illustrates the spatial distribution of seismic events in the study region.

\section{Result and Discussion}
\label{sec:results}
\begin{figure}[htbp]
  \centering
  \includegraphics[width=0.99\textwidth]{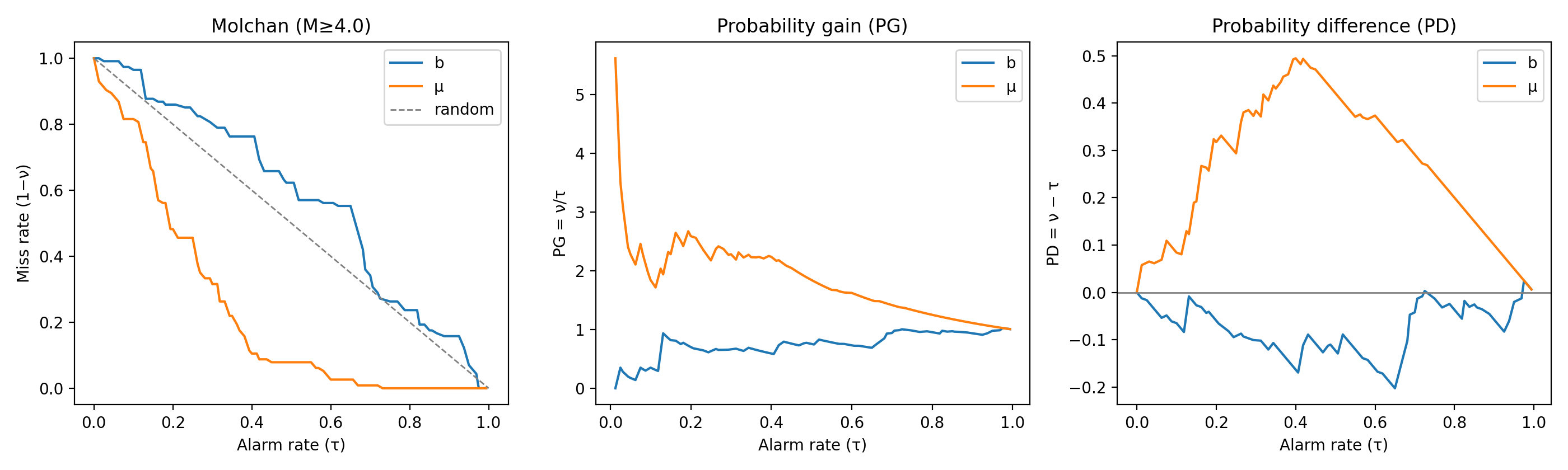} \\
  \vspace{1em}
  \includegraphics[width=0.99\textwidth]{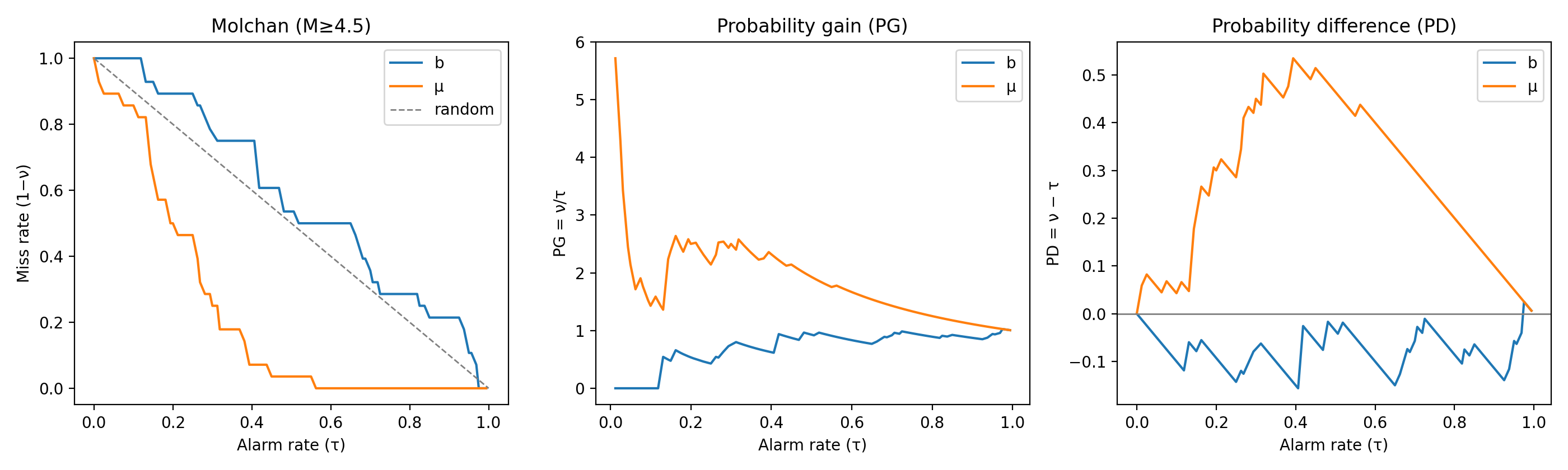}
  \caption{Molchan error curves, probability gain (PG), and probability difference (PD) for earthquakes with {$M \ge 4.0$} and {$M \ge 4.5$}, illustrating the comparative forecasting performance of the {$b$}-value and background seismicity rate {$\mu$}.}
  \label{fig:molchan_curves}
\end{figure}

This research used Molchan diagrams, Probability Gain (\text{PG}), and Probability Difference (\text{PD}) to compare the performance of the {$b$}-value and background seismicity rate ({$\mu$}) in forecasting future earthquakes with {$M \ge 4.0$} and {$M \ge 4.5$} (Figure~\ref{fig:molchan_curves}). Molchan error diagrams serve as the primary diagnostic tool for assessing the performance of alarm-based forecasts against a random spatial reference model. Across nearly the entire range of alarm rates $\tau$, the curve for the background seismicity rate {$\mu$} consistently falls well below the diagonal. This indicates that cells with elevated {$\mu$} capture a disproportionately large share of target earthquakes while occupying only a modest fraction of the study area. Such behavior is fully consistent with physical expectations: {$\mu$} reflects long-term tectonic loading and the persistence of fault activity, and its spatial concentration along active structures naturally makes it a strong predictor of future seismicity \cite{wiemer2002, zechar2010}. The resulting probability gain reaches 5--6 at $\tau < 0.10$, demonstrating that a small subset of high-{$\mu$} cells hosts earthquakes at rates several times higher than chance; an expression of the strong spatial clustering characteristic of the Alborz active fault system. 
The {$b$}-value behaves quite differently at these lower magnitude thresholds. For both {$M \ge 4.0$} and {$M \ge 4.5$}, the {$b$}-value curve lies close to or above the diagonal, and its probability gain remains near unity. In other words, alarming low-{$b$} cells does not meaningfully outperform a random forecast for these magnitudes. This should not be interpreted as a failure of the {$b$}-value concept. Rather, it reflects well-known limitations in {$b$}-value estimation: the metric is most informative when target events are relatively large and sparse, and when the magnitude range lies comfortably above the completeness threshold \cite{schorlemmer2005}. Below roughly {$M_w$}~5.3, spatial instability in {$b$}-value estimation and residual completeness issues tend to obscure the stress-related signal. The negative probability difference values further confirm that, at these magnitudes, {$b$}-value alarms cover more area than they successfully identify as seismically active.

\begin{figure}[htbp]
  \centering
  \includegraphics[width=0.99\textwidth]{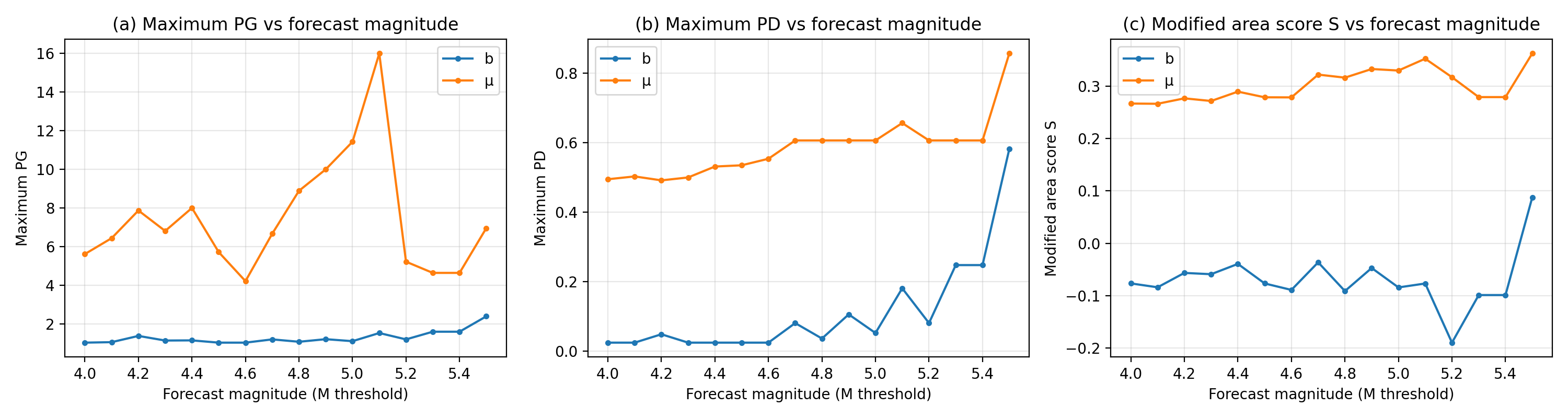}
  \caption{Forecasting performance of spatial {$b$}-value and background seismicity rate ({$\mu$}) for 2025--2029 based on PG, PD, and $S$ analyses.}
  \label{fig:perf_curves}
\end{figure}

Figure~\ref{fig:perf_curves} provides the clearest insight into the magnitude-dependent behavior of both predictors. By examining the maximum probability gain, maximum probability difference, and modified area skill score $S$ as continuous functions of magnitude, it reveals patterns that cannot be captured by a single alarm map or summary statistic. 
The probability gain results show a steady increase in {$b$}-value \text{PG} from $\sim$1.0 at {$M_w$}~4.0 to $\sim$2.5 at {$M_w$}~5.5. This trend is physically intuitive: as magnitude increases, target events become more strongly governed by long-term stress heterogeneity, and low-{$b$} anomalies become more reliable indicators of where larger events are likely to nucleate. The {$\mu$}-based \text{PG} remains higher overall, reflecting its role as the stronger individual predictor, although the spike near {$M \approx 5.1$} likely reflects a cluster of events that happened to fall within high-{$\mu$} cells during the evaluation period.
The probability difference panel reinforces this interpretation. The {$b$}-value \text{PD} rises monotonically from near zero at {$M_w$}~4.0 to $\sim$0.6 at {$M_w$}~5.5, confirming that {$b$}-value heterogeneity becomes increasingly informative with magnitude. In contrast, {$\mu$} shows a relatively stable \text{PD} ($\sim$0.50--0.65) across all thresholds, consistent with its magnitude-independent representation of tectonic loading. The convergence of {$b$}-value and {$\mu$} \text{PD} values at {$M \ge 5.5$} is particularly important: at magnitudes capable of causing structural damage, the two predictors become comparably informative and therefore highly complementary.
The modified area skill score $S$ completes the picture. The {$b$}-value $S$ is negative at low magnitudes, crosses zero near {$M \approx 5.3$}, and becomes modestly positive at {$M \ge 5.5$}. This zero-crossing provides a natural operational threshold: below {$M_w$}~5.3, {$b$}-value information degrades forecast performance, whereas above this threshold it becomes a meaningful and physically grounded contributor. The {$\mu$}-based $S$ remains positive across all magnitudes, reaffirming its role as a stable, magnitude-agnostic spatial prior.

\begin{figure}[htbp]
  \centering
\includegraphics[width=0.99\textwidth]{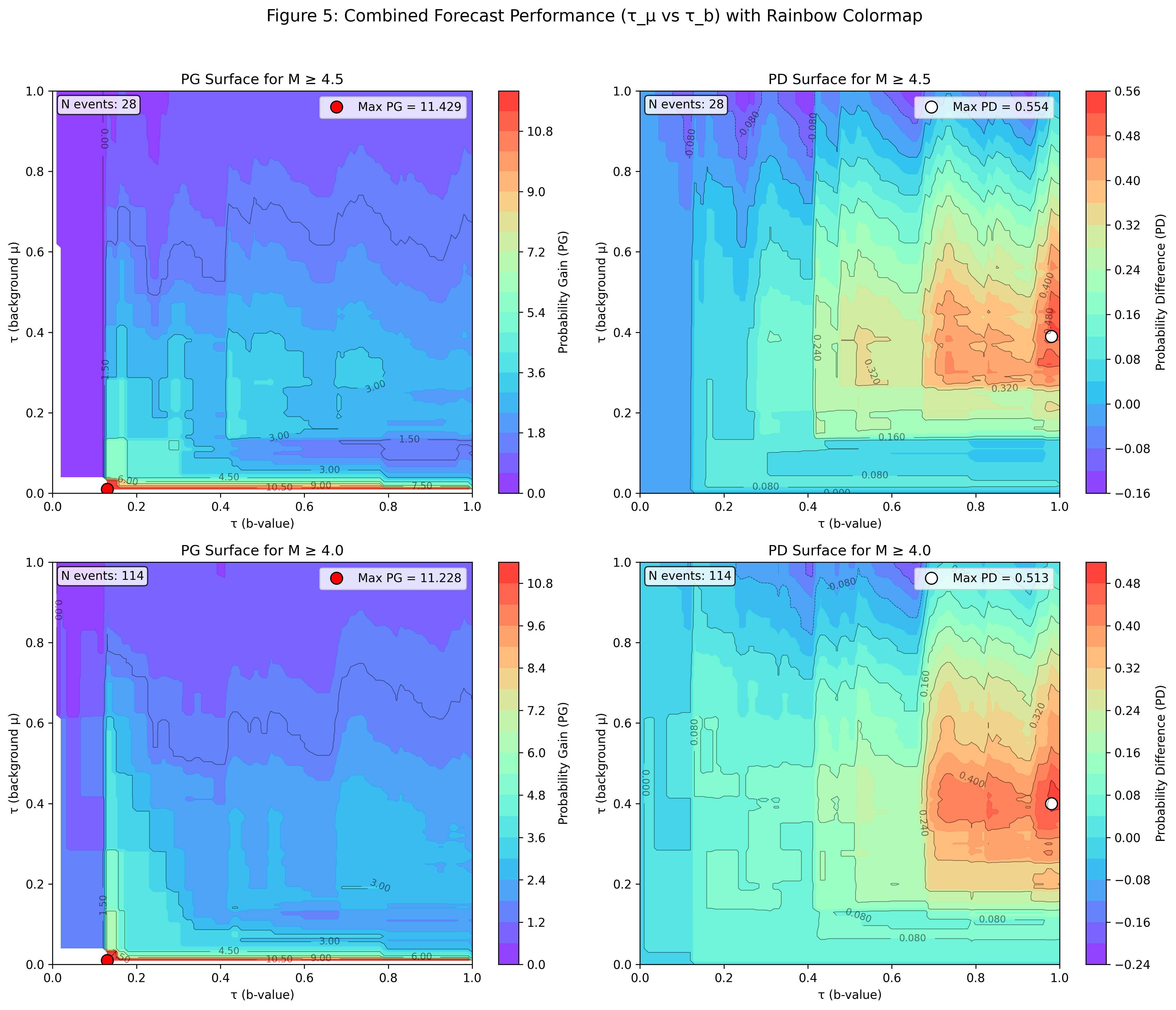}
  \caption{Forecast performance across the $\tau_b$--$\tau_{\mu}$ parameter space for {$M \ge 4.5$} and {$M \ge 4.0$} based on PG and PD metrics, highlighting different sensitivities of the two measures and the dependence of optimal model performance on threshold parameters.}
  \label{fig:parameter_space}
\end{figure}

\begin{figure}[htbp]
  \centering
\includegraphics[width=0.99\textwidth]{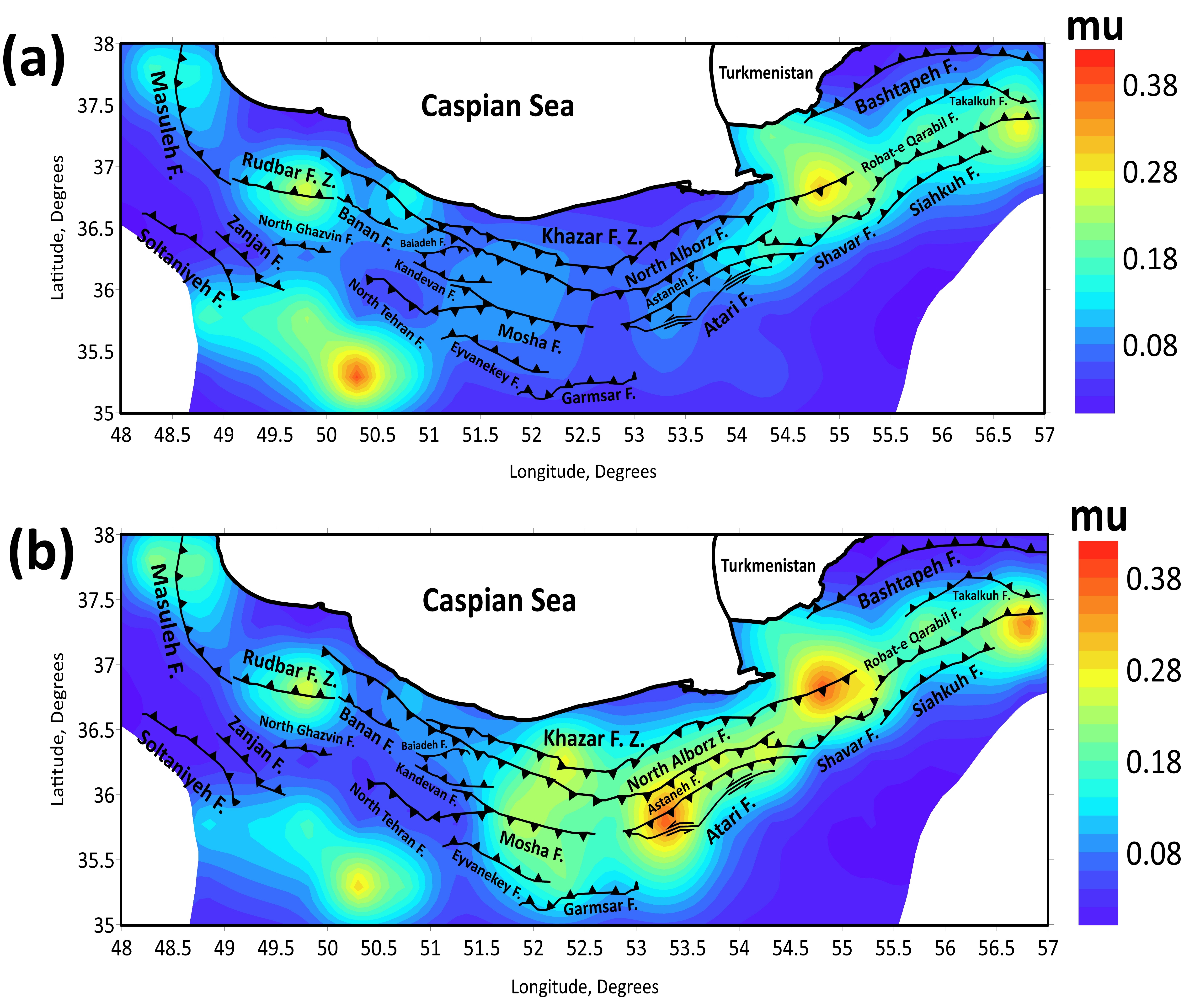}
  \caption{Spatial distribution of the background seismicity rate across the Alborz region for the periods (a) 2006--2015 and (b) 2015--2024. The maps illustrate temporal variations in background seismic activity derived from the ETAS framework.}
  \label{fig:background_rates}
\end{figure}

\begin{figure}[htbp]
  \centering
\includegraphics[width=0.99\textwidth]{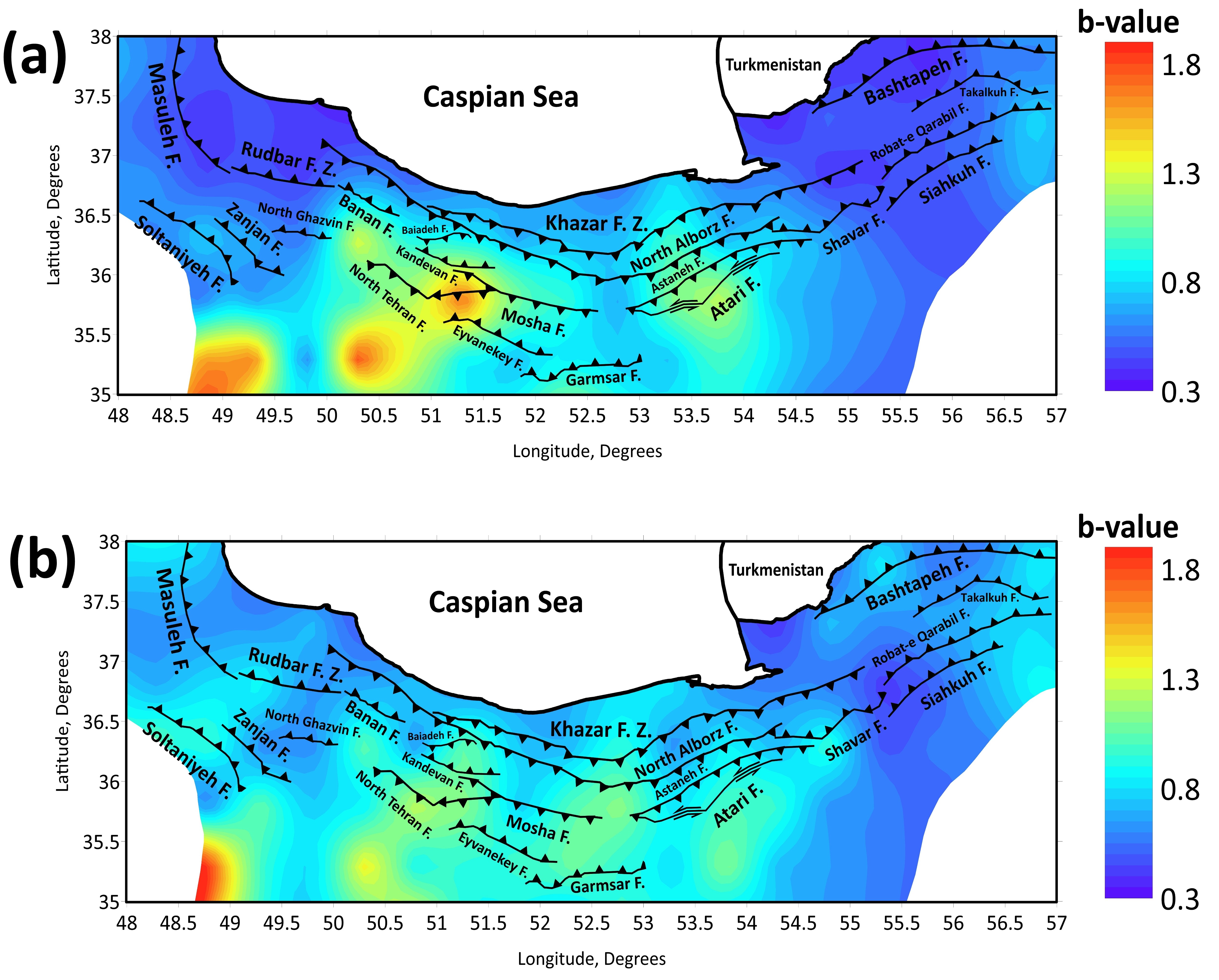}
  \caption{Spatial distribution of {$b$}-values for the periods (a) 2006--2015 and (b) 2015--2024.}
  \label{fig:b_values}
\end{figure}

\begin{figure}[htbp]
  \centering
\includegraphics[width=0.99\textwidth]{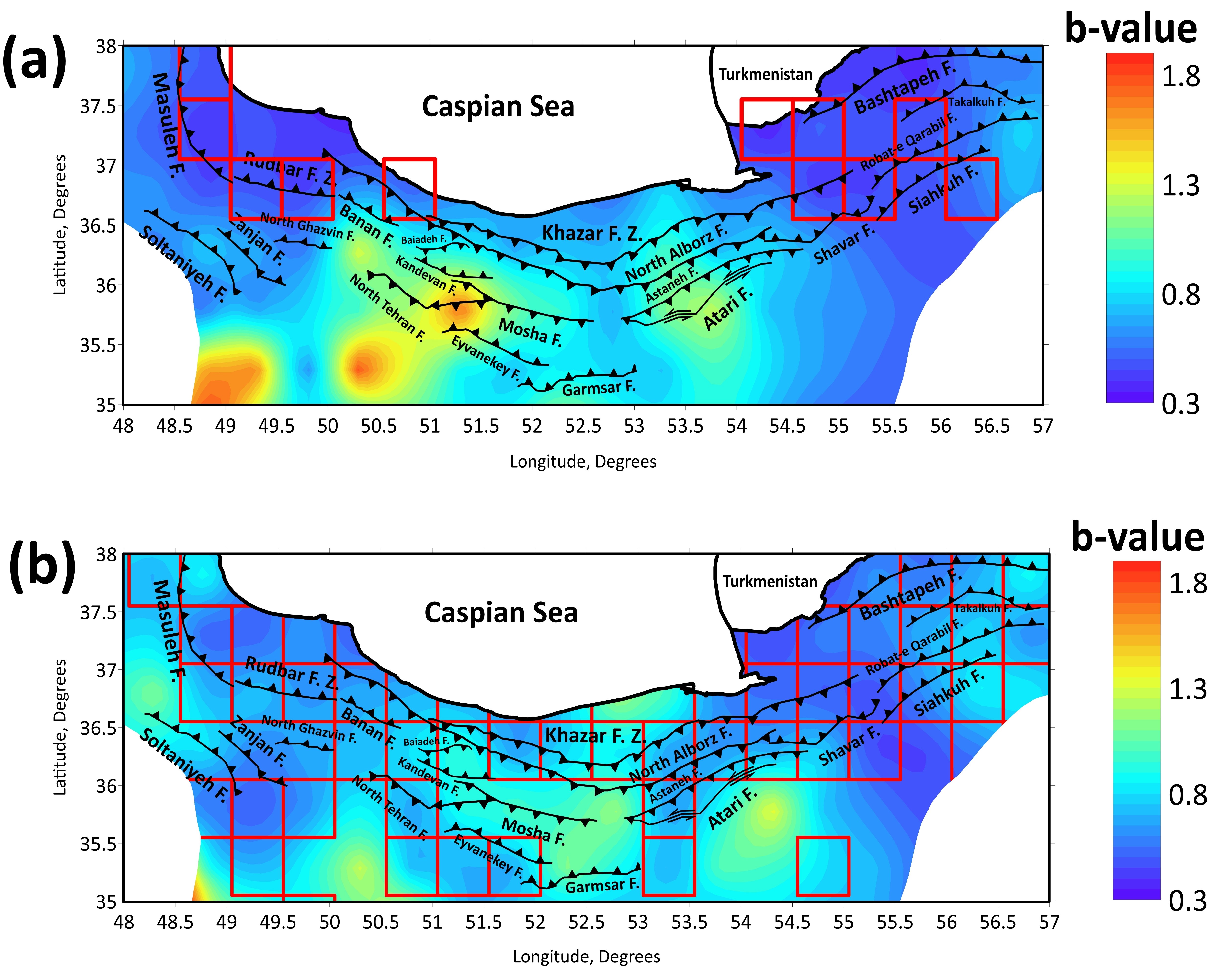}
  \caption{(a) Alarm areas predicted for the 2015--2024 period. (b) Forecasted alarm locations for the 2025--2029 interval.}
  \label{fig:alarms_forecast}
\end{figure}

Figure~\ref{fig:parameter_space} explores how forecast skill evolves when the {$b$}-value and {$\mu$} thresholds are varied jointly, presenting two-dimensional \text{PG} and \text{PD} surfaces across the full parameter space of combined alarm rates ($\tau_b, \tau_{\mu}$) for {$M \ge 4.5$} and {$M \ge 4.0$}. The \text{PG} surfaces confirm what one might expect from the individual results: the highest probability gains are achieved when both thresholds are simultaneously tightened, concentrating alarms in a very small fraction of the study area ($\tau_b < 0.15, \tau_{\mu} \approx 0$, bottom-left corner of the parameter space). This is encouraging from a physical standpoint: it means that the intersection of low-{$b$} and high-{$\mu$} cells defines a genuinely high-hazard zone, not merely a region that happens to be large. 
The \text{PD} surfaces are arguably more useful for operational purposes, precisely because \text{PD} explicitly penalizes both missed events and unnecessary alarms. The \text{PD} extending across $\tau_b = 0.5$--1.0 and $\tau_{\mu} = 0.3$--0.6 indicates that the integrated forecast performs well not just at a single optimal point, but across a wide swath of the parameter space. This robustness means that the forecast conclusions are not hostage to precise threshold calibration.
Looking at the background seismicity rate maps (Figure~\ref{fig:background_rates}), the observation is that seismic activity in the Alborz is far from uniform. During 2006--2015 (Figure~\ref{fig:background_rates}a), the highest {$\mu$} values are found in the western Alborz, clustering around the Rudbar Fault Zone and the Soltaniyeh Fault, with a particularly pronounced hotspot near $49.5^\circ\text{E}, 35.5^\circ\text{N}$. This concentration is hardly surprising given the long shadow cast by the 1990 {$M_w$}~7.4 Rudbar earthquake: even decades after a major rupture, background seismicity tends to remain elevated in its vicinity, reflecting ongoing stress redistribution and possibly fluid-driven reactivation along subsidiary structures \cite{hainzl2005}. 
What is more striking, however, is what happens in the following decade. By 2015--2024 (Figure~\ref{fig:background_rates}b), the seismicity landscape reorganizes itself in a geographically meaningful way. A new zone of elevated {$\mu$} emerges in the central-eastern Alborz, concentrated around the Astaneh and Atari faults ($53^\circ\text{--}54^\circ\text{E}$). The western concentration does not disappear, but it weakens relative to the new activity centers further east. This eastward migration of background seismicity likely reflects genuine changes in the regional stress field, possibly driven by progressive interseismic loading along the North Alborz fault system or by fluid migration at depth, mechanisms that have been documented in comparable compressional tectonic settings \cite{hainzl2005}.
Figure~\ref{fig:b_values} provides a distinct yet complementary characterization of the physical state of the Alborz crust relative to other seismic indicators. Regions where the crust is being squeezed harder tend to produce proportionally more large earthquakes relative to small ones, depressing the {$b$}-value \cite{scholz2015, schorlemmer2005}. During 2006--2015 (Figure~\ref{fig:b_values}a), the most stressed portions of the Alborz, as inferred from low {$b$}-values ($b < 0.8$), are concentrated in the western sector near the Soltaniyeh Fault and across the North Tehran and Eyvanekey fault zones. These are structurally complex zones where multiple fault strands interact under the broad NNE-directed compression imposed by Arabia--Eurasia convergence. The central Alborz shows intermediate {$b$}-values, and the eastern sector near the Bashtapeh and Siahkuh faults appears relatively relaxed, with higher {$b$}-values suggesting lower stress accumulation in that period. 
By 2015--2024 (Figure~\ref{fig:b_values}b), the western stress anomaly largely persists, but something new develops in the central Alborz: a broader low-{$b$} zone expands to encompass the Mosha and Kandevan fault systems. This spatial expansion of the stress anomaly suggests progressive strain accumulation across a broader portion of the range, a pattern consistent with geodetic evidence for distributed shortening along the Alborz \cite{vernant2004}. The eastern Alborz continues to exhibit moderately high {$b$}-values through this period, suggesting that it has not yet accumulated the level of stress loading seen further west.
One particularly important observation emerges when the {$\mu$} and {$b$}-value maps are viewed side by side: they do not mirror each other. Regions of high background seismicity do not systematically coincide with regions of low {$b$}-value, and vice versa. This spatial decoupling is not a methodological inconsistency; it is precisely what one would expect if these two parameters are capturing different physical aspects of the seismogenic system. The {$b$}-value reflects the current stress state of the crust, while {$\mu$} integrates the long-term structural capacity of a region to generate earthquakes. \cite{zhang2025} similarly found no spatial correlation between {$\mu$} and the {$b$}-value in Yunnan Province. They interpreted this lack of correlation as evidence that the two parameters provide complementary rather than overlapping information about the seismotectonic state. The same interpretation is equally applicable to the Alborz region.
Figure~\ref{fig:alarms_forecast} illustrates the spatial {$b$}-value distribution across the Alborz region for the 2015--2024 training period, together with the alarmed grid cells identified by applying the \text{PD}-optimized threshold. Panel (a) presents the {$b$}-value map with alarmed cells overlaid as red-bordered squares, and panel (b) translates this into a forward earthquake forecast for 2025--2029. 
The {$b$}-value map for 2015--2024 reveals a spatially heterogeneous stress field, with the most pronounced low-{$b$} anomalies ($b < 0.8$) concentrated in the western Alborz near the Soltaniyeh and Rudbar fault systems, secondary anomalies in the central Alborz around the Banan and Baladeh faults, and additional low-{$b$} patches in the northeastern sector encompassing the Bashtapeh, Robat-e Qarabil, and Siahkuh fault systems. Application of the \text{PD}-optimized threshold identifies five spatially compact alarmed clusters corresponding to these fault zones, reflecting the selective nature of the low-{$b$} criterion in targeting fault segments with the most compelling evidence of stress concentration. 
The forward forecast for 2025--2029 (Figure~\ref{fig:alarms_forecast}b) exhibits a substantially expanded alarm footprint relative to the training period. Four principal high-risk zones are identified: 
\begin{enumerate}
  \item the western Alborz, incorporating the Masulieh, Rudbar, Zanjan, and Soltaniyeh fault systems;
  \item an extensive central Alborz band encompassing the North Tehran, Mosha, Kandevan, Baladeh, and Khazar fault systems;
  \item the eastern sector around the Astaneh and North Alborz faults ($\sim$52.5$^\circ$--54$^\circ\text{E}$), forming a spatial bridge between the central and northeastern clusters; and
  \item the northeastern Alborz across the Bashtapeh, Robat-e Qarabil, and Siahkuh fault systems.
\end{enumerate}
This spatial expansion reflects the broader low-{$b$} anomaly characterizing the 2015--2024 map and is physically consistent with geodetically observed distributed NNE-directed crustal shortening along the Alborz \cite{vernant2004}.

\section{Conclusion}
\label{sec:conclusion}
The background seismicity rate {$\mu$} emerges as a consistently reliable spatial predictor across all tested magnitude thresholds ({$M \ge 4.0$} and {$M \ge 4.5$}). Its strong performance reflects the direct physical connection between {$\mu$} and long-term tectonic loading, fault architecture, and the persistence of seismic activity along structurally controlled zones. The stability of {$\mu$} across magnitude thresholds further supports its value as a robust spatial prior in seismic hazard assessment. 
The {$b$}-value shows a clear and physically meaningful dependence on magnitude. At lower thresholds ({$M < 5.3$}), spatial variability in {$b$}-value estimates and their proximity to the completeness limit reduce their forecasting capability, which is reflected in negative skill scores and probability gains close to unity. Once magnitudes exceed approximately {$M_w$}~5.3, the skill score becomes positive and {$b$}-value anomalies begin to serve as informative indicators of crustal stress concentration and potential nucleation zones for larger earthquakes. This magnitude-dependent transition provides a practical criterion for when {$b$}-value information can be reliably incorporated into probabilistic forecasting frameworks. 
A spatial comparison of {$b$}-value and {$\mu$} distributions reveals only limited correlation, indicating that the two parameters capture fundamentally different physical aspects of earthquake generation. The {$b$}-value reflects the present differential stress state of the crust, whereas {$\mu$} represents the long-term structural capacity of fault systems to sustain seismic activity. This spatial decoupling aligns with observations from other compressional tectonic settings and highlights the complementary, rather than redundant, nature of the two predictors. 
When combined within a unified forecasting framework, the {$b$}-value and {$\mu$} outperform either predictor used independently. The prospective 2025--2029 spatial forecast, constructed from the 2015--2024 {$b$}-value map using the \text{PD}-optimized threshold, identifies four principal alarmed zones. Taken together, the integration of stress-sensitive and rate-based indicators within a single probabilistic model provides a more physically comprehensive and operationally effective approach to spatial earthquake forecasting. The 2025--2029 forecast offers a testable, time-bounded contribution to seismic hazard assessment in the Alborz and establishes a methodological foundation for future real-time and operational forecasting efforts in tectonically active regions.
\section*{Data Availability}
The earthquake catalog used in this study was obtained from the Iranian Seismological Center (IRSC), Institute of Geophysics, University of Tehran (\url{https://irsc.ut.ac.ir/}). According to the data-use agreement required by the Institute of Geophysics, redistribution of the catalog and any processed or derivative datasets is not permitted. Therefore, the declustered catalog and derived seismicity parameters cannot be made publicly available. These materials may be shared privately upon reasonable request, provided such sharing complies with IRSC restrictions.
\section*{Acknowledgment}
We gratefully acknowledge the Iranian Seismological Center (IRSC) for providing the comprehensive earthquake catalog used in this study, which was instrumental in conducting the spatiotemporal analysis of seismicity.
\section*{Declarations}
\subsection*{Conflict of interest}
The authors declare that they have no known competing financial interests or personal relationships that could have appeared to influence the work reported in this paper.

\bibliography{mybibfile}

\end{document}